\documentclass[12pt,a4paper]{article}%
\usepackage{amsmath,amssymb,amsfonts}
\usepackage{graphicx,epsfig}
\usepackage{color}
\usepackage{amsmath}
\usepackage{amsfonts}
\usepackage{amssymb}
\usepackage{float}
\usepackage{cancel}
\usepackage{caption2}
\usepackage{graphicx}
\usepackage{hyperref}%
\numberwithin{equation}{section} 
\setcaptionmargin{1cm}
\begin{document}
\begin{titlepage}
\begin{flushright}
\end{flushright}
\vskip 1.0cm
\begin{center}
{\Large \bf Higgs doublet as a Goldstone boson in
perturbative\\[5mm]
extensions of the Standard Model}
\vskip 1.0cm {\large Brando Bellazzini$^a$,\ Stefan Pokorski$^b$,\ Vyacheslav S.~Rychkov$^{a,b}$,\\[5mm] Alvise Varagnolo$^{a,c}$} \\[1cm]
{\it $^a$ Scuola Normale Superiore and INFN, Piazza dei Cavalieri 7, I-56126 Pisa, Italy} \\[5mm]
{\it $^b$ Institute of Theoretical Physics, Warsaw University, Hoza 69, 00-681 Warsaw, Poland} \\[5mm]
{\it $^c$ Dip. di Fisica, Univ. di Roma La Sapienza,
P.le A. Moro, 2, I-00185 Rome, Italy}\\[5mm]
\vskip 1.0cm \abstract{We investigate the idea of the Higgs doublet
as a pseudo-Goldstone boson in perturbative extensions of the
Standard Model, motivated by the desire to ameliorate its hierarchy
problem without conflict with the electroweak precision data. Two
realistic supersymmetric models with global $SU(3)$ symmetry are
proposed, one for large and another for small values of $\tan\beta$.
The two models demonstrate two different mechanisms for EWSB and the
Higgs mass generation. Their experimental signatures are quite
different. Our constructions show that a pseudo-Goldstone Higgs
doublet in perturbative extensions is just as plausible as in
non-perturbative ones.}
\end{center}
\end{titlepage}

\section{Introduction}

The idea of the Higgs doublet as a pseudo-Goldstone boson of some extended
global symmetry has been proposed to ameliorate the hierarchy problem of the
Standard Model (SM) \cite{early}. Usually, it is linked to a new strongly
interacting sector, responsible for spontaneous breaking of the global
symmetry \cite{contino},\cite{cont1},\cite{cont2}. There are interesting
signatures of this idea, among others related to the unitarization procedure
in WW scattering \cite{Rattazzi},\cite{Falkowski:2007iv}. However, there are
strong constraints on the scale $f$ of the spontaneous breaking 
of global symmetry of the strong sector. Low values of $f$,
say $f\lesssim500$ GeV, cannot be easily reconciled with electroweak precision
tests and B-physics data \cite{extended}, while larger $f$ reintroduces the
hierarchy problem with the required finetuning growing as\footnote{Here and
throughout the paper $v$ is the electroweak scale in $v\simeq174$ GeV
normalization.} $\left(  f/v\right)  ^{2}$. So, in practice, models of this
kind do not avoid certain tension.

There is some room for the idea of the Higgs doublet as a pseudo-Goldstone
boson in \textit{perturbative} extensions of the SM as well, with global
symmetry broken in the perturbative regime. In general, one may expect such
perturbative models to avoid excessive finetuning in the ElectroWeak Symmetry
Breaking (EWSB) sector with no conflict with the electroweak precision data,
generic for non-perturbative models. This possibility has been discussed in
non-supersymmetric \cite{extended},\cite{csaki} and supersymmetric
\cite{Birkedal:2004xi},\cite{Chankowski:2004mq},\cite{berezhiani}%
,\cite{Roy:2005hg},\cite{Falkowski:2006qq},\cite{strumia} models, however for
various reasons those models are not fully satisfactory. In the present paper
we explore it further in supersymmetric (SUSY) models with extended global
symmetry of the Higgs sector. 
We discuss two models which differ in various respects and illustrate various
aspects of the general approach. As global symmetry we
take $SU(3)$, the minimal one that can give Higgs doublet as a Goldstone boson
in SUSY.

The first model (Model I) remains perturbative up to the GUT scale. The global
symmetry and the electroweak symmetry are broken by radiative corrections to
the mass parameters, generated by a large Yukawa coupling, similarly to the
Minimal Supersymmetric Standard Model (MSSM) (for earlier attempts, see
\cite{berezhiani}). Stabilization of the global symmetry breaking scale $f$
can be achieved by quartic scalar coupling in large $\tan\beta$ regime. 
The model relies on the ``double protection" mechanism, 
where the interplay between supersymmetry and an approximate 
global symmetry forbids the quadratic higgs term to receive a 
large logarithmic contribution from the  UV cutoff 
$\Lambda\sim M_{\text{GUT}}$ which is actually replaced 
by the scale of global symmetry breaking $f$.  
Thus one may hope to get, for the same values of stop mass, 
much less finetuning than in the MSSM.
The values $f\gtrsim2$ TeV minimize finetuning of the model,
while at the same time allowing the physical Higgs boson mass above the
experimental bound of $115$ GeV. Phenomenology of the Higgs sector of the
model is very similar to the decoupling regime of the MSSM. In particular, for
$f\gtrsim2$ TeV the WW scattering is unitarized almost completely by the
lightest Higgs boson. The model is however distinguished by the presence of a
relatively light doubly-charged Higgsino.

In our second example (Model II) supersymmetry provides a consistent framework for
stabilizing the minimum of the global symmetry breaking. 
It is a supersymmetric version of
the approach to the EWSB proposed in Ref. \cite{extended}, with the breaking
driven by a tadpole of the $SU(2)\times U(1)$ singlet component of the full
scalar multiplet. An interesting point about this mechanism is that this
tadpole, while being\textit{ linear} in the fundamental field, generates the
Higgs \textit{quartic} when the $\sigma$-model structure is taken into
account. This quartic dominates the usual D-term quartic at low $\tan\beta$,
so that the physical Higgs mass is determined by the soft SUSY breaking terms
only. Modell II has a very different phenomenology with respect to MSSM since 
it allows for low $f$. However, it needs an UV completion 
at a scale $O(20$ TeV), where the SUSY 
model becomes strongly interacting.

In both models finetuning is $O(10\%).$

\section{Model I}

In the MSSM, the lightest Higgs boson mass is determined by the effective
quartic coupling, which depends logarithmically on the stop mass. Large
$\tan\beta$ is then favored, to minimize the value of the stop mass consistent
with the experimental bound $m_{h}>115$ GeV, and the finetuning in the Higgs
potential. The latter is proportional to $m_{\tilde{t}}^{2}{}\log\Lambda$ and
for $\Lambda\sim M_{\text{GUT}}$ remains, unfortunately, of order of $1\%$.

The model we propose in this section retains the MSSM correlation between the
stop mass and the Higgs boson mass, thus also requiring large $\tan\beta$ for
reasonable values of $m_{\tilde{t}}$. However, it is based on the idea of the
double protection of the Higgs potential \cite{Chankowski:2004mq}%
,\cite{berezhiani} and gives, for the same values of $m_{\tilde{t}}$, factor
$10$ less finetuning than MSSM.

We begin with the effective model below certain scale $F$ based on the
symmetry $SU(2)_{L}\times U(1)_{y}$, where $SU(2)_{L}$ is gauged subgroup of a global $SU(3)$. 
Its UV completion (above $F)$ can be similar to that of \cite{berezhiani}, and we
will return to it below. The $SU(3)$-symmetric Higgs sector consists of a
triplet $\mathcal{H}_{d}$ and an antitriplet $\mathcal{H}_{u}$, while the
chiral fermion multiplets of the top sector are the triplet $\Psi=(Q,T)^{T}$
and quark singlets $t^{c}$ and $T^{c}.$

Under $SU(2)_{L}\times U(1)_{y}$ the triplets split into doublets $H_{u,d}$
and singlets $S_{u,d}$:\footnote{It's useful to keep in mind that our doublet
$H_{u}$ is related to the MSSM doublet $H_{u}$ via $H_{u\text{.our}%
}=\varepsilon\cdot H_{u\text{,MSSM}}$.}%
\[
\mathcal{H}_{d}^{T}=(H_{d}^{T},S_{d}),\quad\mathcal{H}_{u}=(H_{u}%
,S_{u}).\label{split}%
\]
We look for a model in which the global $SU(3)$ is spontaneously broken by
vacuum expectation values (VEVs) aligned so that $SU(2)_{L}\times U(1)_{y}$
gauge symmetry remains unbroken:%
\[
\mathcal{H}_{d}^{T}=(0,0,f_{d}),\quad\mathcal{H}_{u}=(0,0,f_{u}),
\]
and $\tan\beta\equiv f_{u}/f_{d}$ is large. We shall assume that the soft term
mass scale $m_{\text{soft}}\sim f\ll F$. The minimum of the $SU(3)$-symmetric
scalar potential at large $\tan\beta$ generically requires negative
$\mathcal{H}_{u}$ mass squared, which leads to runaway directions, unless there
is a stabilization mechanism. Stabilization by D-terms of some, e.g. $U(1)$, gauge
interactions is not a satisfactory mechanism \cite{Chankowski:2004mq}, while
stabilization by the quartic coupling $\left\vert \mathcal{H}_{u}\right\vert
^{4}$ is constrained by the holomorphicity of the superpotential.

With two Higgs triplet chiral superfields, the minimal field content leading
to stabilization by the quartic consists of two symmetric tensors
$\mathcal{Z}_{1}$ and $\mathcal{Z}_{2}$,%
\[
\mathcal{Z}_{i}=\left(
\begin{array}
[c]{c|c}%
T_{i} & H_{i}/\sqrt{2}\\\hline
H_{i}^{T}/\sqrt{2} & z_{i}%
\end{array}
\right)  .
\]
Here the $T_{i}$'s are $SU(2)$ triplets with $Y_{1,2}=\pm1$, $H_{i}$'s are
doublets with $Y_{1,2}=\pm1/2$, and $z_{i}$'s are singlets. The superpotential
of our model reads (in the following we anticipate large $\tan\beta$ solution)%
\begin{equation}
W=\lambda\mathcal{Z}_{2}\mathcal{H}_{u}\mathcal{H}_{u}+\mu\mathcal{H}%
_{u}\mathcal{H}_{d}+\mu_{Z}\mathcal{Z}_{1}\mathcal{Z}_{2}+y\mathcal{H}_{u}%
\Psi\,t^{c}+mT^{c}T. \label{W1}%
\end{equation}
The last term breaks the global $SU(3)$ explicitly. It can originate from a UV
completion as in \cite{berezhiani}. The scalar potential reads\footnote{The
omitted soft terms $\mathcal{Z}_{1}\mathcal{Z}_{2}$ and $\mathcal{Z}%
_{2}\mathcal{H}_{u}\mathcal{H}_{d}$ will in general be produced by running
from the GUT scale; we have checked that their typical generated values are
small and do not affect the dynamics.}%
\begin{align}
V  &  =|\lambda\mathcal{H}_{u}\mathcal{H}_{u}+\mu_{Z}\mathcal{Z}_{1}%
|^{2}+|2\lambda\mathcal{Z}_{2}\mathcal{H}_{u}+\mu\mathcal{H}_{d}|^{2}+\mu
_{Z}^{2}|\mathcal{Z}_{2}|^{2}+\mu^{2}|\mathcal{H}_{u}|^{2}+V_{\text{soft}%
}\,,\nonumber\\
V_{\text{soft}}  &  =m_{d}^{2}|\mathcal{H}_{d}|^{2}+m_{u}^{2}|\mathcal{H}%
_{u}|^{2}+m_{Z1}^{2}|\mathcal{Z}_{1}|^{2}+m_{Z2}^{2}|\mathcal{Z}_{2}%
|^{2}-(m_{3}^{2}\mathcal{H}_{d}\mathcal{H}_{u}+\text{H.c.})\,. \label{V}%
\end{align}

The soft terms in (\ref{V}) depend on their initial values at the GUT scale
and on the renormalization group (RG) running in the $SU(3)$-symmetric theory.
We expect that the stop contribution will drive $m_{u}^{2}$ to negative values
(while $m_{d}^{2},m_{Z1,Z2}^{2}>0$), and global $SU(3)$ is spontaneously
broken. Minimizing the potential for small $m_{3}^{2}$ (see Appendix
\ref{tech1} for the running of $m_{3}^{2}$) and assuming $SU(3)$ to be broken
in the $SU(2)$ singlet direction, we get%
\begin{align}
&  \langle |\mathcal{H}_{u}|^{2} \rangle \equiv f_{u}^{2}\simeq-\frac{\mu_{u}^{2}}{2\lambda_{\text{eff}%
}^{2}}\label{Su}\\
&  \frac{\langle |\mathcal{H}_{u}|\rangle}{\langle|\mathcal{H}_{d}|\rangle}\equiv\tan\beta\simeq\frac{\tilde{\mu}^{2}+m_{d}^{2}%
}{m_{3}^{2}}\gg1\nonumber\\
&  \langle |\mathcal{Z}_{1}|\rangle \equiv f_{Z1}\simeq-\frac{\lambda\mu_{Z}f_{u}^{2}}{\mu_{Z}^{2}%
+m_{Z1}^{2}}\nonumber\\
&  \langle |\mathcal{Z}_{2}|\rangle \equiv f_{Z2}\simeq-\frac{2\lambda\mu f_{u}f_{d}}{\mu_{Z}^{2}%
+m_{Z2}^{2}+4\lambda^{2}f_{u}^{2}}\nonumber
\end{align}
where%
\begin{align*}
\mu_{u}^{2}  &  =m_{u}^{2}+\mu^{2}<0\text{ (by assumption)}\\
\lambda_{\text{eff}}^{2}  &  =\lambda^{2}\frac{m_{Z1}^{2}}{\mu_{Z}^{2}%
+m_{Z1}^{2}}\\
\tilde{\mu}^{2}  &  =\mu^{2}\frac{m_{Z2}^{2}+\mu_{Z}^{2}}{\mu_{Z}^{2}%
+m_{Z2}^{2}+4\lambda^{2}f_{u}^{2}}%
\end{align*}
The $SU(3)$ is broken dominantly by $f_{u}$ and $f_{Z1}$, with $f_{d}$ and
$f_{Z2}$ suppressed by large $\tan\beta$. Relative contribution of $f_{Z1}$
versus $f_{u}$ decreases for smaller $\lambda$ and $\mu_{Z}$. The maximal
value of $\lambda$ at the Fermi scale is constrained by the requirement of
remaining perturbative up to the GUT scale (see Appendix \ref{tech1} for the
discussion of $\lambda$ running); we choose $\lambda=0.2$ in the following.
The mass parameter $m_{u}^{2}$ gets $SU(3)$-symmetric negative contributions
proportional to the Yukawa coupling $y$ and the coupling $\lambda$ (see
Appendix \ref{tech1}). In the following we will discuss the constraints on the
parameter range following from the demand of no excessive finetuning in the
potential for the $SU(3)$ breaking.

Spontaneous global $SU(3)$ breaking leads to five Goldstone bosons: an
$SU(2)_{L}$ doublet $H$ and a real singlet $\eta$. The $H$ plays the role of
the SM Higgs doublet. The singlet $\eta$ will not play any role in the
following discussion; we will comment on its parametrization and physical
effects below. For large $\tan\beta$ the Goldstones reside to a good
approximation in the $\mathcal{H}_{u}$ and $\mathcal{Z}_{1}$. Up to terms of
higher order in $H$, we have the following parametrization for the Goldstone bosons:%
\begin{align}
H_{d}  &  \simeq\alpha_{d}H,\quad H_{u}\simeq\alpha_{u}H^{\dagger}\nonumber\\
H_{1}  &  \simeq\alpha_{Z1}H^{\dagger},\quad H_{2}\simeq\alpha_{Z2}%
H\label{param}\\
T_{1}  &  \simeq\frac{f_{Z1}}{f^{2}}H^{\dagger}H^{\dagger},\quad T_{2}%
\simeq\frac{f_{Z2}}{f^{2}}H\,H\nonumber\\
S_{u,d}  &  \simeq f_{u,d}\text{,\quad}z_{1,2}\simeq f_{Z1,Z2}\text{,}%
\end{align}
where%
\begin{align*}
\alpha_{u,d}  &  =f_{u,d}/f,\quad\alpha_{Zi}=\sqrt{2}f_{Zi}/f,\\
f^{2}  &  =f_{u}^{2}+f_{d}^{2}+2f_{Z1}^{2}+2f_{Z2}^{2}\,.
\end{align*}
In this parametrization $H$ has canonical kinetic term. As we will see later,
experimental limit on the Higgs mass requires $f_{Z1,Z2}\ll f_{u}\sim f.$

The global $SU(3)$ is explicitly broken by the last term in the superpotential
(\ref{W1}) and by the D-terms of $SU(2)\times U(1)$. Both terms contribute to
the potential for the Goldstone boson $H$:%
\[
V=\delta m_{H}^{2}|H_{u}|^{2}+(\lambda_{0}+\delta\lambda)|H_{u}|^{4}+\ldots,
\]
where the $\delta m_{H}^{2}$ and $\delta\lambda$ are obtained from the
one-loop effective potential and $\lambda_{0}$ comes from the D-terms. We
first discuss the effective potential contribution as it is responsible for
the VEV of $H$ and the EWSB by the top-stop loops. Diagonalizing the top mass
matrix, for large $\tan\beta$ we find (we introduce dimensionless coupling
$\tilde{y}$, $m=\tilde{y}\langle|S_{u}|\rangle$):%
\begin{align*}
m_{t}  &  =y_{t}\langle |H_{u}|\rangle ,\quad y_{t}\equiv\frac{y\tilde{y}}{\sqrt{y^{2}%
+\tilde{y}^{2}}},\\
m_{T}  &  =\langle |S_{u}|\rangle \sqrt{y^{2}+\tilde{y}^{2}}.
\end{align*}
For the couplings $y$ and $\lambda$ to remain perturbative up to the GUT scale
(see Appendix), we need $y\lesssim1.2$. Since $y_{t}\simeq1$ (for
$\langle|H_{u}|\rangle\simeq v$) we get $\tilde{y}\gtrsim1.8$ and $m_{T}\gtrsim2.2\langle|S_{u}|\rangle$,
somewhat stronger than the theoretical lower bound $m_{T}=2y_{t}\langle|S_{u}|\rangle$
realized for $y\simeq\tilde{y}\simeq\sqrt{2}$.

To realize the double protection mechanism, we assume that soft stop masses
are $SU(3)$-symmetric at the scale $F$. To compute the effective potential, we
assume for simplicity that these masses are universal:
\[
m_{Q}^{2}(|\tilde{Q}|^{2}+|\tilde{T}^{c}|^{2}+|\tilde{t}^{c}|^{2})\text{, }%
\]
and we also neglect the possible left-right stop mixing. As in ref.
\cite{berezhiani} we get the following result:%
\begin{equation}
\delta m_{H}^{2}=-\frac{3}{8\pi^{2}}y_{t}^{2}\left[  m_{Q}^{2}\ln\left(
1+\frac{m_{T}^{2}}{m_{Q}^{2}}\right)  +m_{T}^{2}\ln\left(  1+\frac{m_{Q}^{2}%
}{m_{T}^{2}}\right)  \right]  +\Delta\text{,} \label{mass}%
\end{equation}
where%
\begin{equation}
\Delta\supset\frac{3g_{2}^{2}M_{2}^{2}+g_{y}^{2}M_{y}^{2}}{8\pi^{2}}\ln
\frac{F}{M_{\text{soft}}}, \label{gtomass}%
\end{equation}
the contribution due to $SU(2)\times U(1)_{y}$ gauginos with soft masses
$M_{2}$,$M_{y}$. At the same time the dominant contribution to $\delta\lambda$
is given by\footnote{The formula given in \cite{berezhiani} contains an extra
$+3/2$ term in square brackets. Our formula is correct provided that
$\delta\lambda$ is defined as the coefficient in the Higgs mass correction
formula, Eq.\ (\ref{Higgsmass}).}%
\begin{equation}
\delta\lambda\simeq\frac{3}{16\pi^{2}}y_{t}^{4}\left[  \ln\frac{m_{Q}^{2}%
}{m_{t}^{2}(1+x)}-2x\ln(1+1/x)\right]  ,\quad x=m_{Q}^{2}/m_{T}^{2}.
\label{quartic}%
\end{equation}
Notice that this correction is \textit{smaller} than the corresponding MSSM
correction for the same value of the stop mass, which is due to the negative
contribution of the heavy $T$ quark. In the $m_{T}\gg m_{Q}$ limit, which will
turn out to be relevant below, we recover the standard MSSM equations, with
the important difference that the scale of the logarithm in (\ref{mass}) is
given by $m_{T}$ instead of $M_{\text{GUT}}$.

The D-term potential reads:%
\[
V_{D}=\frac{g^{2}+g_{y}^{2}}{8}\left(  |H_{u}|^{2}-|H_{d}|^{2}+|H_{1}%
|^{2}-|H_{2}|^{2}\right)  ^{2}=\frac{g^{2}+g_{y}^{2}}{8}(\alpha_{u}^{2}%
-\alpha_{d}^{2}+\alpha_{Z1}^{2}-\alpha_{Z2}^{2})^{2}|H|^{4}.
\]
For the Higgs boson mass we get the following result:%
\begin{equation}
m_{h}^{2}\simeq\left(  1-v^{2}/f^{2}\right)  \left[  M_{Z}^{2}(\alpha_{u}%
^{2}-\alpha_{d}^{2}+\alpha_{Z1}^{2}-\alpha_{Z2}^{2})^{2}+4\delta
\!\lambda\,\alpha_{u}^{4}\,v^{2}\right]  . \label{Higgsmass}%
\end{equation}
The overall suppression factor is due to the $\sigma$-model correction to the
wavefunction normalization of the Higgs doublet; it can be derived by keeping
track of terms higher order in $H$ which were omitted in (\ref{param}). 
Considering the large $\tan\beta$ suppression resulting 
in $\alpha_{d}$ and $\alpha_{Z2}$ going to zero, we see
that for a given $\delta\lambda$ the Higgs boson mass is maximized for
\begin{equation}
\tan\beta\rightarrow\infty,\quad\alpha_{Z1}\rightarrow0,\quad f\rightarrow
\infty, \label{optimal}%
\end{equation}
\begin{equation}
m_{h}^{\text{max}}=\left(  M_{Z}^{2}+4\delta\!\lambda\,v^{2}\right)  ^{1/2}.
\label{mhmass}%
\end{equation}
Expanding in the small negative corrections appearing 
when these parameters deviate from
their optimal values,  (\ref{Higgsmass}) can be numerically parametrized as follows
\begin{equation}
m_{h}\simeq m_{h}^{\text{max}}-1\text{ GeV}\left[  \left(  \frac{12}{\tan
\beta}\right)  ^{2}+\left(  \frac{1.3\text{ TeV}}{f}\right)  ^{2}+\left(
\frac{\alpha_{Z1}}{0.15}\right)  ^{2}\right]  , \label{mcorr}%
\end{equation}
where the first and the second corrections come from finite values of $\tan\beta$ and $f$ respectively, 
and the third from a nonzero $\alpha_{Z1}$.
Since $m_{h}^{\text{max}}$ cannot be much above 115 GeV 
without a significant increase in finetuning (see
Figs \ref{mhiggs-fig}, \ref{Delta1} below), we should
not allow the total loss in (\ref{mcorr}) to exceed $1\div2$ GeV, which
implies obvious constraints on the relevant parameters.

We now discuss the results for the Higgs boson mass and estimate the level of
finetuning. In Fig. \ref{mhiggs-fig} we plot the Higgs boson mass
(\ref{mhmass}) (i.e. without negative corrections given in (\ref{mcorr})) as a
function of $m_{Q}$ and $m_{T}$, using $\delta\lambda$ from eq. (\ref{quartic}%
) with $m_{t}=172$ GeV. Similarly to the MSSM, the Higgs boson mass increases
with $m_{Q}$. For a fixed $m_{Q}$, the correction is maximized in the
$m_{T}\gg m_{Q}$ limit.%
\begin{figure}
[ptb]
\begin{center}
\includegraphics[
height=2.4076in,
width=2.5002in
]%
{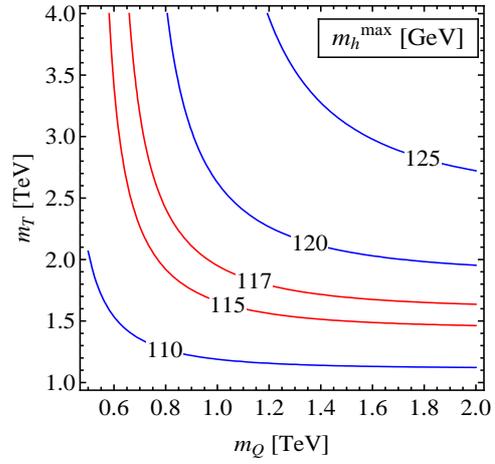}%
\caption{The maximal Higgs boson mass (\ref{mhmass}) as a function of $m_{Q}$
and $m_{T}$, see the text.}%
\label{mhiggs-fig}%
\end{center}
\end{figure}

In Fig. \ref{Delta1} we plot the finetuning in the Higgs mass term
$m^{2}|H|^{2}$ which is needed to compensate the top-stop contribution
(\ref{mass}):%
\begin{equation}
\text{FT}_{1}=\frac{\delta m_{H}^{2}|_{\Delta=0}}{  m_{h}^{2}/2}  \,. \label{FT1}%
\end{equation}%
\begin{figure}
[ptb]
\begin{center}
\includegraphics[
height=2.4422in,
width=2.5356in
]%
{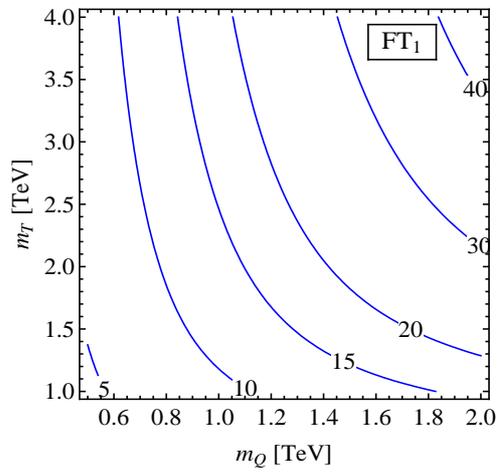}%
\caption{Finetuning (\ref{FT1}) in the Higgs mass parameter needed to
compensate for the top-stop loop contribution (\ref{mass}).}%
\label{Delta1}%
\end{center}
\end{figure}
This finetuning increases quadratically with $m_{Q}$, but grows only
logarithmically with $m_{T}$. Comparing Fig. \ref{mhiggs-fig} with Fig.
\ref{Delta1}, we see that $m_{Q}\sim800$ GeV and $m_{T}\gtrsim2.5$ TeV give
$m_{h}>115$ GeV with about 10\% finetuning (FT$_{1}=10$). Soft stop masses as
small as $m_{Q}\sim600$ GeV are possible provided that $m_{T}$ is raised up to
$4$ TeV.

\begin{figure}
[ptb]
\begin{center}
\includegraphics[
height=2.207in,
width=2.2917in
]%
{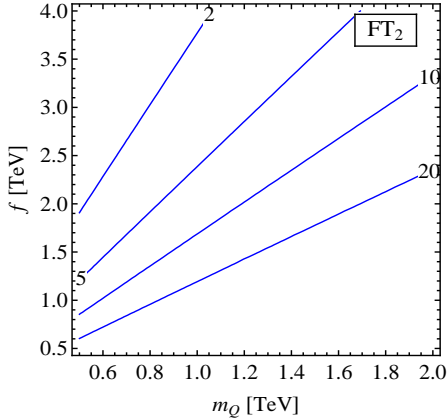}%
\caption{Finetuning (\ref{FT2}) in the $SU(3)$-symmetric mass parameter
$m_{u}^{2}$, see the text.}%
\label{Delta2}%
\end{center}
\end{figure}

Another source of finetuning in Model I is the $SU(3)$-symmetric top-stop
contribution to the $m_{u}^{2}$ parameter of the scalar potential (\ref{V}),
which by (\ref{Su}) should not exceed $2\lambda^{2}f_{u}^{2}.$ The
corresponding finetuning parameter%
\begin{equation}
\text{FT}_{2}=\delta m_{u,stop}^{2}/(2\lambda^{2}f^{2}) \label{FT2}%
\end{equation}
is plotted in the $m_{Q}-f$ plane in Fig. \ref{Delta2}, where we assume
$y\simeq1,$ $\lambda=0.2$. We see that this finetuning is less than $20$\%
(FT$_{2}<5$) for $m_{Q}\sim800$ GeV and $f\gtrsim2$ TeV, which however
translates into $m_{T}$ above $4$ TeV.%

In general raising $f$ (and $m_{T}\simeq2f)$ we eliminate FT$_{2}$ while
FT$_{1}$ grows only logarithmically. Unfortunately, in this limit the heavy
top quark becomes undiscoverable at the LHC, and the scalar spectrum of the
model resembles the standard MSSM at large $\tan\beta$ in the decoupling
limit. In this case the only significant difference from the MSSM is the
presence in the low-energy spectrum of states described 
by the tensors $\mathcal{Z}_{1}$ and $\mathcal{Z}_{2}$, 
i.e. triplets $T_{i}$, doublets $H_{i}$ and  singlets $z_{i}$.  
The triplets contain doubly  ($\tilde{T}^{++}_{1}$ or $\tilde{T}^{--}_{2}$)  
and singly ($\tilde{T}^{+}_{1}$ or $\tilde{T}^{-}_{2}$) charged
and neutral ($\tilde{T}^{0}_{1}$ or $\tilde{T}^{0}_{2}$) higgsinos  and their scalar partners. 
Doublets $H_i$ have the same composition as the
Higgs chiral superfields $H_{u,d}$.  
All those fields have common supersymmetric mass parameter $\mu_Z$.
It enters in the equation (\ref{Su}) for the VEV\ of $z_{1}$
and is constrained by the requirement that no large negative $\alpha_{Z1}%
$-correction to the Higgs boson mass should be present in Eq.\ (\ref{mcorr}).
Assuming that all scalar soft masses are of the same order:%
\[
M_{\text{SUSY}}\sim m_{u}\sim\sqrt{2}\lambda f\sim0.3f\text{,}%
\]
$\mu_{Z}$ is bounded by%
\[
\mu_{Z}\lesssim\alpha_{Z1}m_{\text{soft}}\lesssim f/20\text{.}%
\]
This estimate, while being subject to significant uncertainty, does indicate
that the masses of new fermions, in particular of the doubly charged Higgsinos, are expected to be below $200\div300$ GeV even for $f$ as high as $2$ TeV. We shall return to the phenomenological issues
at the end of this section.

Finally, we need to comment on several other issues which are important for
the consistency of our model. First, we note that the model can be UV
completed as in Ref. \cite{berezhiani}, with the gauge group $SU(3)\times
U(1)_{x}$ broken to the electroweak $SU(2)\times U(1)_{y}$ at the scale $F$.
An extra pair of triplets $\Phi_{U,D}$ is responsible for this breaking, so
that the full global symmetry of the scalar potential is $SU(3)\times
SU(3)$.\ The $SU(3)$-breaking term $mT^{c}T$ in the Model I superpotential
(\ref{W1}) originates naturally from an $SU(3)$-symmetric term
\begin{equation}
y_{1}\Phi_{U}\Psi T^{c},\quad y_{1}\sim m/F=\tilde{y}f/F, \label{int1}%
\end{equation}
of the UV completed superpotential. As explained in \cite{berezhiani}, the
soft mass terms of the $\Phi_{U}$ and $\Phi_{D}$ fields have to be very nearly
universal at $F$, since their difference $m_{D}^{2}-m_{U}^{2}$ contributes to
the mass term of the Higgs doublet via D-terms. Even assuming that these
masses are universal at the GUT scale, superpotential interaction (\ref{int1})
will contribute to the running of $m_{U}^{2}$, so that at $F$ the masses will
be split by%
\[
m_{D}^{2}-m_{U}^{2}\simeq\frac{3y_{1}^{2}}{8\pi^{2}}(m_{Q}^{2}+m_{T^{c}}%
^{2})\ln\frac{M_{\text{GUT}}}{F}%
\]
This contribution must be $\lesssim v^{2}$ which can be achieved by choosing
$F\gtrsim10f$ as can be seen from the second eq.\ in (\ref{int1}).

On the other hand the $F$ scale cannot be too high because of the gaugino
contributions to the Higgs mass, Eq.\ (\ref{gtomass}).

For completeness we have to say a few words about the 5th Goldstone boson
$\eta$, a gauge singlet axion, which appears in addition to the Higgs doublet
when $SU(3)$ is broken spontaneously to $SU(2)$, as already mentioned above.
This Goldstone is associated with a global $U(1)$ under which the gauge
singlet components have charges $S_{u}(+1/2),S_{d}(-1/2),z_{1}(+1),z_{2}(-1)$,
equal to the hypercharge of the upper components of the same $SU(3)$
multiplets; it resides mostly in the phase of $S_{u}$ whose VEV dominates the
spontaneous symmetry breaking:%
\[
S_{u}\simeq f_{u}\exp(i\eta/\sqrt{2}f_{u})\,.
\]
The $\eta$ does not get mass from the $SU(3)$-breaking terms which we so far
considered, since they preserve the above $U(1)$; it can however get mass if
we add a small $SU(3)$-breaking tadpole%
\begin{equation}
\Delta V_{\text{soft}}=-m_{S}^{3}S_{u}+\text{H.c.,} \label{tadpole}%
\end{equation}
which gives $m_{\eta}^{2}=m_{S}^{3}/f.$ This term breaks $S_{u}\rightarrow
-S_{u}$ symmetry and can be generated radiatively by adding to the
superpotential a small term $\Delta W=m^{\prime}Tt^{c}$ breaking the same
symmetry:%
\[
m_{S}^{3}\simeq-\frac{3y}{2\pi^{2}}m^{\prime}m_{Q}^{2}\ln\frac{M_{\text{GUT}}%
}{F}.
\]
It should be noticed however that the discussed axion, even if exactly
massless, would not be in conflict with experiment \cite{berezhiani}, since it
couples very weakly to the ordinary matter (such a coupling could for example
proceed via mixing with heavy fermions which are needed to implement the
$SU(3)$ symmetry in the first and second generations).

Our last comment concerns the impact of the triplets $T_{1},$ $T_{2}$ on the
precision electroweak observables. According to Eq.\ (\ref{param}) they get
VEVs, $T_{i}^{11}\simeq f_{Zi}\frac{v^{2}}{f^{2}}$. Since according to our
discussion below Eq.\ (\ref{mcorr}) we need $f_{Z1}<0.1f$ (and $f_{Z2}$ is
even smaller), the contribution of the triplets to the $\rho$ parameter is
sufficiently suppressed:
\[
\delta\rho=-2\frac{(T_{1}^{11})^{2}}{v^{2}},\quad|\delta\rho|<10^{-4}\left(
\frac{1.7\text{ TeV}}{f}\right)  ^{2}.
\]

We conclude that Model I is a fully consistent example of a supersymmetric
model with a Higgs doublet as a Goldstone boson of extended global symmetry,
perturbative up to the GUT scale and with large $\tan\beta$. Its phenomenology
is similar to phenomenology of the MSSM in the decoupling limit, but
finetuning in the Higgs potential is diminished by at least a factor $10$
compared to MSSM. The lightest Higgs mass is expected to be just above $115$
GeV, with stop around $800$ GeV, and the new top quark above $3\div4$ TeV,
probably unreachable at the LHC. However, stabilization of the $SU(3)$
breaking potential requires new states. The extended scalar sector of Model I
is unlikely to manifest itself at the LHC, since these particles are expected
to be quite heavy (apart from the decoupled axion $\eta$), while their
couplings to $WW$ and $t\bar{t}$ are suppressed due to large $\tan\beta$ and
$f/v$ ratio. 

Among new fermions, there are those with the same
quantum numbers of MSSM chargino and neutralino states. 
However, the mixing between MSSM states and these new fermions 
is small because $f_{Zi}\ll f$  and the mass eigenstates remain almost
MSSM-like. 
The new mass eigenstates, 
including $\tilde{T}^{\pm\pm}$ and $\tilde z_{i}$'s, are almost degenerate
in mass, with masses $m\approx \mu_Z \approx 200$~GeV. 
The details of the mass spectrum depend on the details of the
mixing.  
Thus we expect new light fermions doubling the 
MSSM states, but the best chance to see a trace of the $SU(3)$ structure is
probably the doubly charged Higgsino. At the LHC, this
double-chargino $\tilde{T}^{\pm\pm}$ will be pair-produced via the Drell-Yan
process. It will most likely undergo a chain decay into $\tilde{T}^{\pm}$ 
and $W$ and finally into the lightest neutralino and two same sign $W$ bosons. 
Because of an approximate degeneracy of the mass spectra the decay
products are likely to be off-shell. The double-chargino is likely to be long-lived and the decay vertex may be displaced.
In any case, we expect events with two
opposite-hemisphere pairs of same-sign leptons and missing transverse energy
in the final state. We are not aware of experimental studies for signals of
this type\footnote{In \cite{Demir:2008wt} 
doubly charged Higgsinos with a significant coupling to leptons 
were considered, so that a dominant decay mode is into
slepton-lepton pairs, with the slepton subsequently decaying into a
lepton plus neutralino. This gives rise to a practically background-free
same-sign, same-flavor lepton pair and missing energy signature.
Such couplings in our model would violate the lepton number conservation and are 
by no means necessary (if allowed at all). 
Our case is definitely more challenging experimentally.}.
By analogy with Drell-Yan chargino-neutralino searches, we can
optimistically estimate the LHC integrated luminosity required for the
discovery of $\tilde{T}^{\pm\pm}$ to be $O(100)$ fb$^{-1}$. 
The fermionic neutral singlets $\tilde{z}_{i}$'s may also be phenomenologically interesting. Depending on the details of the mass
spectrum one of them may be the LSP and a potential candidate for dark matter particle.
Our model provides then a concrete example of a spectrum going beyond the MSSM spectrum. More detailed phenomenological
study of such and similar spectra are of experimental interest but beyond the scope of this paper.

\section{Model II}

It has been emphasized in \cite{extended} (following an earlier observation in
\cite{contino}), that in models with extended global symmetry there exists
also a mechanism for EWSB and the Higgs boson mass generation based on a
tadpole contribution of an $SU(2)\times U(1)$ singlet component of the full
scalar multiplet. This mechanism necessarily requires a small value of $f$, to
minimize finetuning in the Higgs potential. In turn, this implies large
quartic coupling and low UV cutoff. On the other hand, high enough Higgs boson
mass can be produced even for moderate $\tan\beta$.

The simplest model achieving stabilization of $SU(3)$ breaking at small $f$
includes two Higgs triplets $\mathcal{H}_{d}$ and $\mathcal{H}_{u}$, same as
in Model I, and an $SU(3)$ singlet $N$, with the superpotential%
\[
W=\kappa N\mathcal{H}_{u}\mathcal{H}_{d}.
\]
As we will see, it generically leads to low values of $\tan\beta$. We must
have $\kappa\leq2$ for the Landau pole to be above $\Lambda_{\text{UV}}=20-30$
TeV; we will choose $\kappa=2$ in what follows. All RG runnings below are
considered from $\Lambda_{\text{UV}}$ down to the Fermi scale, $\log
\frac{\Lambda_{\text{UV}}}{M_{\text{SUSY}}}\simeq3$ for $M_{\text{SUSY}}\sim1$ TeV.

The scalar potential reads:%
\begin{align}
V  &  =\kappa^{2}[|\mathcal{H}_{u}\mathcal{H}_{d}|^{2}+|N|^{2}(|\mathcal{H}%
_{u}|^{2}+|\mathcal{H}_{d}|^{2})]+V_{\text{soft}}\nonumber\\
V_{\text{soft}}  &  =m_{u}^{2}|\mathcal{H}_{u}|^{2}+m_{d}^{2}|\mathcal{H}%
_{d}|^{2}+m_{N}^{2}|N|^{2}-(A^{3}N+m_{3}^{2}\mathcal{H}_{u}\mathcal{H}%
_{d}+\text{H.c.}) \label{V2}%
\end{align}
The masses $m_{u,d}^{2},m_{N}^{2}$ need to be positive to avoid runaway. We
use $m_{3}^{2}$ to break $SU(3)$ spontaneously, while $A^{3}$ will give a VEV
to $N$ and generate an effective $\mu$-term (chargino mass). It is consistent
to assume that all terms which are not included at the tree level remain small
or vanish. E.g. $A^{\prime}NH_{1}H_{2}$ term is generated by gaugino masses,%
\[
\delta A^{\prime}\sim\frac{g^{2}\kappa M_{2}}{16\pi^{2}}\times3\sim
0.02M_{2}\,.
\]
Possible modifications of the model can be obtained by adding $\mu
\mathcal{H}_{u}\mathcal{H}_{d}$ and/or $\kappa F^{2}N$ terms to the
superpotential, as well as $N\mathcal{H}_{u}\mathcal{H}_{d}$ term to
$V_{\text{soft}}$. These modifications lead to models of comparable
\textquotedblleft quality", and we will not consider them.

{}Minimization of the potential (\ref{V2}) in the gauge singlet direction
gives%
\begin{align}
\langle|\mathcal{H}_{u,d}|\rangle  &  \equiv f_{u,d},\quad f^{2}\equiv f_{u}^{2}+f_{d}^{2}=\frac
{m_{3}^{2}-\mu_{u}\mu_{d}}{\kappa^{2}\sin\beta\cos\beta},\quad\tan\beta
\equiv\frac{f_{u}}{f_{d}}=\frac{\mu_{d}}{\mu_{u}},\label{minII}\\
\mu_{u,d}^{2}  &  \equiv m_{u,d}^{2}+\mu^{2},\quad\mu\equiv\kappa f_{N},\quad
\langle|N|\rangle\equiv f_{N}=\frac{A^{3}}{m_{N}^{2}+\kappa^{2}f^{2}},\nonumber
\end{align}
where we continue using notation of section \ref{split} for the components of
$\mathcal{H}_{d}$ and $\mathcal{H}_{u}$.

As in Model I, spontaneous breaking of $SU(3)$ to $SU(2)$ generates five
Goldstone bosons, a doublet $H$ and an axion. $SU(3)$ symmetry must then be
broken also explicitly, to get a potential for $H$, which will break
electroweak symmetry. In this model the presence of the soft term $m_{3}^{2}$ 
results in nonvanishing VEVs $f_{u,d,N}$ breaking the global $SU(3)$.
To avoid any risk
of destabilization of the $SU(3)$ breaking potential by a negative $m_{u}^{2}%
$, we keep the top-stop sector as in MSSM. Thus, $SU(3)$ is explicitly broken
by the top-stop sector. The standard RG running from $\Lambda_{\text{UV}}$
generates a negative contribution to the Higgs mass squared
\begin{equation}
-\delta m_{H}^{2}|H_{u}|^{2}\equiv-\delta m_{H}^{2}\left(  |\mathcal{H}%
_{u}|^{2}-|S_{u}|^{2}\right)  \label{dmH}%
\end{equation}
Another source of the explicit $SU(3)$ breaking is a tadpole contribution
$m_{S}^{3}S_{u}$, which, we assume, is generated by strong dynamics at
$\Lambda_{\text{UV}}$\footnote{Perturbative origin of the tadpole could be
engineered if desired by breaking $S_{u}\rightarrow-S_{u}$ symmetry in the
superpotential, similarly to the generation of (\ref{tadpole}) in Model I.}.
Making use of (\ref{dmH}), we can write the full $SU(3)$ breaking potential
as a function of $S_{u}$:
\begin{equation}
\Delta V=m_{H}^{2}|S_{u}|^{2}-\left(  m_{S}^{3}S_{u}+\text{H.c.}\right)
\text{,} \label{po-br}%
\end{equation}
which has the clear advantage of completely decoupling the $SU(3)$-symmetric
potential minimization and vacuum disalignment. In this parametrization, we
view the $|\mathcal{H}_{u}|^{2}$ contribution in (\ref{dmH}) as a
renormalization of $m_{u}^{2}$ parameter in (\ref{V2}). Minimizing the CP-even
part of (\ref{po-br}), we find the VEV\ of $S_{u}$:
\[
\langle|S_{u}|\rangle=\frac{m_{S}^{3}}{m_{H}^{2}}\text{,}%
\]
where we have to assume that the found minimum satisfies $|S_{u}|<f_{u}$,
otherwise the true minimum will be at $S_{u}\simeq\pm f_{u}$ with no EWSB. On
the other hand, $|S_{u}|<f_{u}$ means vacuum disalignment, with the Higgs VEV%
\begin{equation}
\langle|H_{u}|^{2}\rangle\equiv v_{u}^{2}=f_{u}^{2}-\langle|S_{u}|^{2}\rangle \label{Higgs-vev}%
\end{equation}
and the EWSB scale given by%
\begin{equation}
v^{2}=v_{u}^{2}+v_{d}^{2}=f^{2}\left[  1-\left(  \frac{m_{S}^{3}}{f_{u}%
m_{H}^{2}}\right)  ^{2}\right]  . \label{vev0}%
\end{equation}
We see that $v\ll f$ can be obtained only at the price of finetuning the ratio
$m_{S}^{3}/f_{u}m_{H}^{2}$ to $1$. To illustrate this finetuning more clearly,
we can express (\ref{po-br}) as a function of $|H_{u}|$ by expanding
$S_{u}=\sqrt{|f_{u}|^{2}-|H_{u}|^{2}}$, which is a good
approximation for $v\ll f$:
\begin{equation}
\Delta V\simeq const-(m_{H}^{2}-m_{S}^{3}/f_{u})|H_{u}|^{2}+\lambda|H_{u}%
|^{4}+\ldots,\quad\lambda=\frac{m_{S}^{3}}{4f_{u}^{3}}\,. \label{expan}%
\end{equation}
We now see the origin of finetuning: it appears since we are canceling the
$O(f^{2})$ Higgs mass term with the quadratic term appearing in the expansion
of $\sqrt{f_{u}^{2}-|H_{u}|^{2}}$, taking advantage of the non-linear
structure of the $\sigma$-model.

The finetuning discussed above can be quantified by means of the usual
logarithmic derivative, or by measuring the portion of the uniformly
distributed parameter space satisfying $v\leq174$ GeV; we get \cite{extended}%
\footnote{This finetuning does not increase even if $m_{H}$ and $m_{S}$ are
scaled up, because the Higgs quartic $\lambda$ in (\ref{expan}) and,
correspondingly, the Higgs mass (see below), increases in the same limit,
\textquotedblleft improving naturalness" in the sense of \cite{IDM}.}
\begin{equation}
\text{FT}\simeq\frac{2f^{2}}{v^{2}}\,. \label{FT-II}%
\end{equation}
As a reference value we fix $f=350$ GeV, corresponding to $O(10)\%$
finetuning\footnote{Comparing with \cite{extended}, notice a factor $\sqrt{2}$
difference in normalization of $f$ resulting from the change from real to
complex fields. Our $f=350$ GeV gives the same finetuning as $f\simeq500$ GeV
in \cite{extended}.}.

Other potential sources of finetuning in Model II are related to the RG
running of the $SU(3)$-symmetric potential parameters. According to
Eq.\ (\ref{minII}) to avoid large cancellations in the potential for $SU(3)$
breaking, we must have
\begin{equation}
\mu_{u}\mu_{d}\lesssim\kappa^{2}f^{2}/2 \label{nat0}%
\end{equation}
i.e. effectively
\[
\mu^{2},\,m_{u}^{2},\,m_{d}^{2}\lesssim\kappa^{2}f^{2}/4\,.
\]
As mentioned above, the $|\mathcal{H}_{u}|^{2}$ in Eq.\ (\ref{dmH})
effectively renormalizes $m_{u}^{2};$ naturalness thus requires that
\begin{equation}
\delta m_{H}^{2}=\frac{3}{4\pi^{2}}y_{t}^{2}m_{\tilde{t}}^{2}\log\frac
{\Lambda_{\text{UV}}}{M_{\text{SUSY}}}\sim\frac{m_{\tilde{t}}^{2}}{4\sin
^{2}\beta}\lesssim m_{u}^{2}\lesssim\kappa^{2}f^{2}/4\,. \label{mhII}%
\end{equation}
Thus we get
\begin{equation}
m_{\tilde{t}}\lesssim\sin\beta\,\kappa f\,. \label{mstconstr}%
\end{equation}

Finally, we discuss the Higgs boson mass. The same expansion of the square
root which allows us to finetune $v\ll f$ in (\ref{expan}), also generates a
Higgs quartic $\lambda$. For small $\tan\beta$ this quartic easily dominates
the standard $D$-term quartic; as a result the Higgs boson mass is solely
determined by the soft terms and the coupling $\kappa$. Taking into account the $\sigma$-model
wavefunction suppression and also using the exact expression of the Higgs
potential (\ref{po-br}) instead of expanding in $H_{u}$, we find the Higgs boson mass
\[
m_{h}=\sin\beta(m_{H}/f)v\,.
\]
If we assume that $m_{H}^{2}$ is entirely generated by stop loops we get
\[
m_{h}=\frac{v}{2}\frac{m_{\tilde{t}}}{f}\,,
\]
and thus $m_{\tilde{t}}\simeq\sqrt{2}f$ for $m_{h}\simeq120$ GeV, consistently
with the constraint (\ref{mstconstr}) on $m_{\tilde{t}}$ for low $\tan\beta$.
The model then predicts a light Higgs boson since larger values of
$m_{\tilde{t}}/f$ are inconsistent with the constraint (\ref{mstconstr}).

We will discuss phenomenology of Model II for $f=350$ GeV, $1\leq\tan\beta
\leq2$, and $\kappa=2$.We choose the potential parameters as follows
($t\equiv\tan\beta$):
\begin{align*}
\mu^{2}  &  =m_{u}^{2}=\frac{\kappa^{2}f^{2}/2}{1+t^{2}},\quad\,m_{d}%
^{2}=\frac{2t^{2}-1}{t^{2}+1}\kappa^{2}f^{2}/2\,,\\
m_{3}^{2}  &  =\frac{2\kappa^{2}f^{2}}{t+t^{-1}},\quad m_{N}^{2}=6m_{u}%
^{2},\quad\,A^{3}=(m_{N}^{2}+\kappa^{2}f^{2})\frac{\mu}{\kappa}\,,
\end{align*}
which is consistent with naturalness and produces a minimum of the potential
at the given values of $f$ and $\tan\beta$. We see that Higgsinos are expected
to be light, $\mu\sim100\div200$ GeV, for $f=350$ GeV.

Furthermore, since $f$ is small, phenomenology of the model is strongly
influenced by its non-linear structure. We recall that in the $\sigma$-model
approximation
\begin{equation}
\mathcal{H}_{u}=f_{u}\left(
\begin{array}
[c]{l}%
\frac{H}{|H|}\sin(\frac{|H|}{f})\\
\cos(\frac{|H|}{f})
\end{array}
\right)  \label{Huparam}%
\end{equation}
and similarly for $\mathcal{H}_{d}$ ($|H|=\sqrt{H^{\dagger}H}$). We also can
parametrize the Higgs doublet $H$ nonlinearly:
\begin{equation}
H=\Sigma\left(
\begin{array}
[c]{c}%
0\\
\bar{v}+h/\sqrt{2}%
\end{array}
\right)  , \label{goldst}%
\end{equation}
where $\Sigma=e^{iT^{a}G^{a}/v}$ is the pion field containing the Goldstone
bosons eaten by the W and Z. The fields are canonically normalized. The true
electroweak scale reads $v=f\sin(\bar{v}/f)$. Additional heavy scalar modes
describe deviation from the $\sigma$-model structure (\ref{Huparam}). For
instance, fluctuations in the radial directions can be introduced by replacing%
\begin{equation}
f_{u,d}\rightarrow f_{u,d}+s_{u,d}/\sqrt{2}. \label{radial}%
\end{equation}
We are interested in the couplings of the scalars $h$, $s_{u}$,
$s_{d}$ to the vector boson pairs. By the equivalence theorem, these
couplings can be obtained from the kinetic part of the Lagrangian for
$\mathcal{H}_{u}$ and $\mathcal{H}_{d}$ inserting (\ref{goldst}) into
(\ref{Huparam}). The couplings of $s_{u,d}$ are found using
(\ref{radial}). We get:
\begin{equation}
\mathcal{L}=v^{2}|D_{\mu}\Sigma|^{2}+\sqrt{2}v|D_{\mu}\Sigma|^{2}\left[
\cos(\bar{v}/f)h+\frac{v}{f}(\cos\beta\,s_{u}+\sin\beta\,s_{d})\right]
\end{equation}
We see that the $h$ coupling to pions, and hence to $WW,$ is suppressed by
$\cos(\bar{v}/f)=\left(  1-v^{2}/f^{2}\right)  ^{1/2}$, and $h$ unitarizes
$WW$ amplitude only partially. Unitarization is completed by the exchange of
the heavy scalars. The $s_{u,d}WW$ couplings appear because the radial
directions obtain nonzero projection on the first two components of
$\mathcal{H}_{u,d}$ when expanded around $v\neq0$.

The fields $s_{u}$ and $s_{d}$ are not mass eigenstates. In
the mass matrix, they mix with each other and also with the radial
excitation of $N$. Thus, three heavy mass eigenstates complete the
unitarization of $WW$ scattering. Denoting the mass eigenstates by $S_{i}$,
$i=1,2,3$, e.g. for $f=350$ GeV, $\kappa=2$ and $\tan\beta=2$ we get
$m_{S_{i}}\simeq(290,850,1000)$ GeV$.$ The cubic $WW$-scalar interaction
Lagrangian in this case is given by%
\begin{align}
\mathcal{L}  &  =g_{WWh}^{\text{SM}}[\cos(\bar{v}/f)h+(v/f)c_{i}S_{i}]W_{\mu
}^{2},\label{cubic}\\
c  &  \simeq(0.86,-0.13,-0.5),\quad c_{i}^{2}=1,\nonumber
\end{align}
so that the $WW$ scattering is fully unitarized above $m_{S3}$.

Other important couplings are the $S_{i}\bar{t}t$ couplings as they determine
the production rate of these scalars via the gluon fusion. They originate from
the term
\[
y_{t}\frac{v}{f}\frac{s_{u}}{\sqrt{2}}t\bar{t},\quad y_{t}\equiv
\frac{m_{t}}{v\sin\beta},
\]
which appears similarly to the $s_{u,d}WW$ couplings discussed above.
Due to the $v/f$ coupling suppression, production rates of the heavy scalars
via the gluon fusion, as well as via the vector boson fusion, will be
suppressed by at least one order of magnitude with respect to the
corresponding production rates for the SM Higgs boson of the same mass.
Nevertheless, at least the lightest of these heavy scalars should be quite
easy to discover at the LHC in the gold-plated decays $S_{1}\rightarrow
ZZ\rightarrow4l$.

Apart from the radial modes (\ref{radial}), another interesting heavy mode is
a longitudinal fluctuation orthogonal to the pseudo-Goldstone mode:%
\begin{equation}
H_{u}=\cos\beta\,H_{1},\quad H_{d}=-\sin\beta H_{1}. \label{long}%
\end{equation}
The $SU(2)$ doublet $H_{1}$ does not get a VEV and is decoupled from the
vector boson pairs; it is analogous to the heavy MSSM doublet in the
decoupling limit. It describes a degenerate heavy multiplet ($H^{\pm}%
,H^{0},A^{0})$ of mass%
\[
m_{H1}^{2}\simeq\frac{\mu_{u}\mu_{d}}{\sin\beta\cos\beta},
\]
which can be found substituting (\ref{long}) into the scalar potential. By
(\ref{nat0}), we expect $m_{H1}=O(\kappa f)\sim700$ GeV. The neutral members
of this multiplet couple to $t\bar{t}$ with strength $\cot\beta$ times the SM
Higgs coupling. They will be produced via gluon fusion and will be seen as
narrow resonances decaying into $t\bar{t}$ pairs (total width around $30$
GeV). Using the model-independent analysis of \cite{walker}, we can estimate
that $O(10)$ fb$^{-1}$ of integrated luminosity could be enough for their
discovery at the LHC.

Finally, we discuss the effect of the heavy scalars on the electroweak
observables. As pointed out in Ref.\cite{extended}, the relevant parameter is
the effective Higgs boson mass, which in our case is given by
\[
m_{\text{EWPT}}=m_{h}\left(  \frac{\bar{m}}{m_{h}}\right)  ^{\frac{v^{2}%
}{f^{2}}},\quad\bar{m}=\prod\left(  m_{S_{i}}\right)  ^{c_{i}^{2}},
\]
where the $c_{i}$ are the parameters appearing in the $WW$-scalar interaction
Lagrangian (\ref{cubic}). In our numerical example we get $\bar{m}=400$ GeV
and $m_{\text{EWPT}}=155$ GeV for $m_{h}=115$ GeV. Thus, $m_{\text{EWPT}}$ is
slightly above the $144$ GeV $95\%$ C.L. limit, but various other
supersymmetric contributions can easily compensate its effect in the $(S,T)$ plane.

\section{Conclusions}

We have presented two realistic supersymmetric models with Higgs doublet as
Goldstone boson of a spontaneously broken extended global symmetry. Model I is
perturbative up to the GUT scale and realizes large $\tan\beta$ scenario,
while Model II requires a rather low UV cut-off ($\sim20$ TeV) and generically
gives low $\tan\beta$. Both models avoid excessive finetuning in the Higgs
potential and are in fact motivated by this requirement. Being perturbative up
to much higher cut-off than so-called \textquotedblleft strongly
interacting\textquotedblright\ models, they do not lead to any serious tension
with precision electroweak data. The two models illustrate two different
mechanisms for EWSB and the Higgs mass generation. Their experimental
signatures are quite different. Clearly, the price for a small finetuning is
some complexity (e.g. compared to the MSSM). Our constructions supplement the
list of previous proposals for ameliorating the supersymmetric little
hierarchy problem. E.g. the Next-to-Minimal Supersymmetric Standard Model
easily solves the little hierarchy if its $SH_{u}H_{d}$ coupling $\lambda$ is
allowed to become strong below the GUT scale (a possibility recently taken to
the extreme in \cite{lsusy}). Its predictions are different from Model II as,
for instance, it does not predict non-linear effects in the scalar couplings.
We will wait and see what experiment tells us. For the moment, the main lesson
of our constructions is that the possibility of the Higgs boson as a Goldstone
boson in perturbative theories looks equally plausible as in non-perturbative
scenarios with low cut-off and actually more predictive.

\section{Acknowledgements}

V.S.R. was supported by the EU under RTN contract MRTN-CT-2004-503369 and ToK
contract MTKD-CT-2005-029466.

\appendix

\section{Technical details on Model I}

\label{tech1}

In this appendix we collect some technical details relevant for Model I. It is
convenient to normalize all charges and couplings with the UV completion into
$SU(3)\times U(1)_{x}$ in mind. For instance $\mathcal{H}_{u}$ has
$SU(3)\times U(1)_{x}$ quantum numbers $3_{1/3}$,
\[
D\mathcal{H}_{u}=\left(  W^{a}T^{a}+1/3B_{x}\right)  \mathcal{H}_{u}\,.
\]
The unbroken generator is $Y=\frac{1}{\sqrt{3}}T^{8}+X$. The $SU(3)$ gauge
coupling $g$ coincides at the scale $F$ with the $SU(2)$ gauge coupling
$g_{2}$. The $U(1)_{y}$ coupling $g_{y}$ at $F$ is:
\[
g_{y}=\frac{gg_{x}}{\sqrt{g^{2}+g_{x}^{2}/3}}\,.
\]
Numerical values of $g$ and $g_{x}$ are $0.65$ and $0.37$ respectively. The RG
equation for $m_{3}^{2}$ in Eq.\ (\ref{V}) valid above $F$ reads
\[
16\pi^{2}\frac{dm_{3}^{2}}{d\log\Lambda}=-\frac{16}{3}g^{2}M\mu+U(1)_{x}%
\text{-gaugino contribution.}%
\]
Here $M$ is the $SU(3)$-symmetric gaugino mass. For the running from the GUT
scale we get
\[
\delta m_{3}^{2}\sim0.4M\mu\,.
\]
From (\ref{Su}), the natural value of $\tan\beta$ is $m_{d}^{2}/\delta
m_{3}^{2}$. We see that $\tan\beta=O(10)$ is naturally allowed, provided that
$m_{d}$ is a factor of a few larger than $M,\mu$. Running from $F$ down
generates only $H_{u}H_{d}$ coefficient which is much smaller since the
running is very short.

Other mass parameters whose running is of interest for the model are
$m_{u}^{2}$ and $m_{Z2}^{2}$. For $\log M_{\text{GUT}}/F\sim30$ one gets
\begin{align*}
\delta m_{u}^{2}  &  =\frac{1}{16\pi^{2}}6y^{2}\left(  m_{Q}^{2}+m_{T^{c}}%
^{2}+m_{u}^{2}\right)  \times30+\frac{1}{16\pi^{2}}16\lambda
^{2}\left(  m_{u}^{2}+m_{Z2}^{2}\right)  \times30\\
\delta m_{Z2}^{2}  &  =\frac{1}{16\pi^{2}}4\lambda^{2}\left(  m_{u}^{2}%
+m_{Z2}^{2}\right)  \times30
\end{align*}
The dominant contribution comes from the term proportional to the Yukawa
coupling. It makes $m_{u}^{2}$ negative and breaks the global $SU(3)$ symmetry radiatively.

Finally, we discuss the perturbativity constraint up to the GUT scale on the
couplings $\lambda$ and $y$. The RG equations read:
\begin{align*}
16\pi^{2}\frac{dy}{d\log\Lambda}  &  =y\left(  7y^{2}+8\lambda^{2}-\frac
{16}{3}(g_{3}^{2}+g^{2})-O(g_{x}^{2})\right) \\
16\pi^{2}\frac{d\lambda}{d\log\Lambda}  &  =\lambda\left(  18\lambda
^{2}+6y^{2}-12g^{2}-\frac{4}{3}g_{x}^{2}\right)
\end{align*}
where $g_{3}$ is the strong coupling constant. One can check that the safe
range for values of the couplings at the scale $F$ is $y\lesssim1.2$,
$\lambda\lesssim0.3.$


\begin{thebibliography}{99}                                                                                               %


\bibitem {early}D.~B.~Kaplan and H.~Georgi, \textquotedblleft SU(2) X U(1)
Breaking By Vacuum Misalignment,\textquotedblright\ Phys.\ Lett.\ B
\textbf{136}, 183 (1984).
D.~B.~Kaplan, H.~Georgi and S.~Dimopoulos, \textquotedblleft Composite Higgs
Scalars,\textquotedblright\ Phys.\ Lett.\ B \textbf{136}, 187 (1984).


\bibitem {contino}K.~Agashe, R.~Contino and A.~Pomarol, \textquotedblleft The
minimal composite Higgs model,\textquotedblright\ Nucl.\ Phys.\ B
\textbf{719}, 165 (2005) [arXiv:hep-ph/0412089].


\bibitem {cont1}R.~Contino, L.~Da Rold and A.~Pomarol, ``Light custodians in
natural composite Higgs models,'' Phys.\ Rev.\ D \textbf{75}, 055014 (2007)
[arXiv:hep-ph/0612048].


\bibitem {cont2}R.~Contino, T.~Kramer, M.~Son and R.~Sundrum,
``Warped/Composite Phenomenology Simplified,'' JHEP \textbf{0705}, 074 (2007)
[arXiv:hep-ph/0612180].


\bibitem {Rattazzi}G.~F.~Giudice, C.~Grojean, A.~Pomarol and R.~Rattazzi,
\textquotedblleft The strongly-interacting light Higgs,\textquotedblright JHEP
\textbf{0706}, 045 (2007),\ arXiv:hep-ph/0703164.


\bibitem {Falkowski:2007iv}A.~Falkowski, S.~Pokorski and J.~P.~Roberts,
\textquotedblleft Modelling strong interactions and longitudinally polarized
vector boson scattering,\textquotedblright\ JHEP \textbf{0712}, 063 (2007)
[arXiv:0705.4653 [hep-ph]].


\bibitem {extended}R.~Barbieri, B.~Bellazzini, V.~S.~Rychkov and A.~Varagnolo,
\textquotedblleft The Higgs boson from an extended symmetry,\textquotedblright%
\ Phys.\ Rev.\ D \textbf{76}, 115008 (2007) arXiv:0706.0432 [hep-ph].


\bibitem {csaki}C.~Csaki, J.~Heinonen, M.~Perelstein and C.~Spethmann, ``A
Weakly Coupled Ultraviolet Completion of the Littlest Higgs with T-parity,''
arXiv:0804.0622 [hep-ph].


\bibitem {Birkedal:2004xi}A.~Birkedal, Z.~Chacko and M.~K.~Gaillard,
\textquotedblleft Little supersymmetry and the supersymmetric little hierarchy
problem,\textquotedblright\ JHEP \textbf{0410}, 036 (2004)
[arXiv:hep-ph/0404197].


\bibitem {Chankowski:2004mq}P.~H.~Chankowski, A.~Falkowski, S.~Pokorski and
J.~Wagner, \textquotedblleft Electroweak symmetry breaking in supersymmetric
models with heavy scalar superpartners,\textquotedblright\ Phys.\ Lett.\ B
\textbf{598}, 252 (2004) [arXiv:hep-ph/0407242].


\bibitem {berezhiani}Z.~Berezhiani, P.~H.~Chankowski, A.~Falkowski and
S.~Pokorski, \textquotedblleft Double protection of the Higgs
potential,\textquotedblright\ Phys.\ Rev.\ Lett.\ \textbf{96}, 031801 (2006)
[arXiv:hep-ph/0509311].


\bibitem {Roy:2005hg}T.~Roy and M.~Schmaltz, \textquotedblleft Naturally heavy
superpartners and a little Higgs,\textquotedblright\ JHEP \textbf{0601}, 149
(2006) [arXiv:hep-ph/0509357].


\bibitem {strumia}C.~Csaki, G.~Marandella, Y.~Shirman and A.~Strumia, ``The
super-little Higgs,'' Phys.\ Rev.\ D \textbf{73}, 035006 (2006)
[arXiv:hep-ph/0510294].


\bibitem {Falkowski:2006qq}A.~Falkowski, S.~Pokorski and M.~Schmaltz,
\textquotedblleft Twin SUSY,\textquotedblright\ Phys.\ Rev.\ D \textbf{74},
035003 (2006) [arXiv:hep-ph/0604066].


\bibitem {IDM}R.~Barbieri, L.~J.~Hall and V.~S.~Rychkov, \textquotedblleft
Improved naturalness with a heavy Higgs: An alternative road to LHC
physics,\textquotedblright\ Phys.\ Rev.\ D \textbf{74}, 015007 (2006)
[arXiv:hep-ph/0603188].


\bibitem {walker}V.~Barger, T.~Han and D.~G.~E.~Walker, \textquotedblleft Top
Quark Pairs at High Invariant Mass - A Model-Independent Discriminator of New
Physics at the LHC,\textquotedblright\ Phys.\ Rev.\ Lett.\ \textbf{100},
031801 (2008) [arXiv:hep-ph/0612016].


\bibitem {lsusy}R.~Barbieri, L.~J.~Hall, Y.~Nomura and V.~S.~Rychkov,
``Supersymmetry without a light Higgs boson,'' Phys.\ Rev.\ D \textbf{75},
035007 (2007) [arXiv:hep-ph/0607332].


\bibitem{Demir:2008wt}D.~A.~Demir, M.~Frank, K.~Huitu, S.~K.~Rai and I.~Turan,
``Signals of Doubly-Charged Higgsinos at the CERN Large Hadron Collider,''
arXiv:0805.4202 [hep-ph].

\end{thebibliography}
\end{document}